\documentclass{emulateapj}

\newcommand{\myemail}{sascha.quanz@astro.phys.ethz.ch}

\slugcomment{Revised version April 13, 2015}

\shorttitle{Confirmation and characterization of the protoplanet HD100546 b}
\shortauthors{Quanz et al.}

\begin{document}


\title{Confirmation and characterization of the protoplanet HD100546 b --- 
Direct evidence for gas giant planet formation at 50 au}

\author{Sascha P. Quanz$^{1,2}$, Adam Amara$^2$, Michael R. Meyer$^2$, Julien H. Girard$^3$, Matthew A. Kenworthy$^4$, and Markus Kasper$^5$}
%\affil{Institute for Astronomy, ETH Zurich, Wolfgang-Pauli-Strasse 27, 8093 Zurich, Switzerland}    
\email{\myemail}

\altaffiltext{1}{Based on observations collected at the European Organisation for Astronomical Research in the Southern Hemisphere, Chile, under program number 091.C-0818(A).}
\altaffiltext{2}{Institute for Astronomy, ETH Zurich, Wolfgang-Pauli-Strasse 27, 8093 Zurich, Switzerland}
\altaffiltext{3}{European Southern Observatory, Alonso de C\'ordova 3107, Vitacura, Cassilla 19001, Santiago, Chile}
\altaffiltext{4}{Sterrewacht Leiden, P.O. Box 9513, Niels Bohrweg 2, 2300 RA Leiden, The Netherlands}  
\altaffiltext{5}{European Southern Observatory, Karl Schwarzschild Strasse, 2, 85748 Garching bei M\"unchen, Germany}

\begin{abstract}
We present the first multi-wavelength, high-contrast imaging study confirming the protoplanet embedded in the disk around the Herbig Ae/Be star HD100546. The object is detected at $L'$ ($\sim 3.8\,\mu m$) and $M'$ ($\sim 4.8\,\mu m$), but not at $K_s$ ($\sim 2.1\,\mu m$), and the emission consists of a point source component surrounded by spatially resolved emission. 
For the point source component we derive apparent magnitudes of $L'=13.92\pm0.10$ mag, $M'=13.33\pm0.16$ mag, and $K_s>15.43\pm0.11$ mag (3$\sigma$ limit), and a separation and position angle of $(0.457\pm0.014)''$ and $(8.4\pm1.4)^\circ$, and $(0.472\pm0.014)''$ and $(9.2\pm1.4)^\circ$ in $L'$ and $M'$, respectively. We demonstrate that the object is co-moving with HD100546 and can reject any (sub-)stellar fore-/background object. Fitting a single temperature blackbody to the observed fluxes of the point source component yields an effective temperature of $T_{eff}=932^{+193}_{-202}$ K and a radius for the emitting area of $R=6.9^{+2.7}_{-2.9}$ R$_{\rm Jupiter}$. The best-fit luminosity is $L=(2.3^{+0.6}_{-0.4})\cdot 10^{-4}\,L_\sun$. We quantitatively compare our findings with predictions from evolutionary and atmospheric models for young, gas giant planets, discuss the possible existence of a warm, circumplanetary disk, and note that the de-projected physical separation from the host star of $(53\pm2)$ au poses a challenge standard planet formation theories. Considering the suspected existence of an additional planet orbiting at $\sim$13--14 au, HD100546 appears to be an unprecedented laboratory to study the formation of multiple gas giant planets empirically. 
\end{abstract}


\keywords{planets and satellites: detection -- planets and satellites: formation -- planets and satellites: gaseous planets -- protoplanetary disks -- planet-disk interactions -- stars: pre-main sequence}



\section{Introduction}
The formation of gas giant planets within dust and gas rich disks surrounding young stars is complex and theoretical models describing the immediate formation process are barely constrained by empirical data. At the moment there are two main theories: the core accretion model based on the physics of mutual collisions and growth of solid bodies followed by accretion of a gaseous envelope \citep[e.g.,][]{pollack1996}, and the gravitational instability model predicting direct, local collapse in the outer regions of circumstellar disks \citep[e.g.,][]{boss2001}. Up to now, these theories are indirectly constrained, e.g., by studying the chemical and physical conditions in circumstellar disks to estimate the global initial conditions for planet formation \citep[e.g.,][]{dutrey2014}, or by studying the composition of extrasolar planets, both from estimates of the bulk density and from atmospheric characterization, to decipher their formation history and evolution \citep[e.g.,][]{marcy2014,konopacky2013}. Furthermore, models describing the luminosity evolution of young gas giant planets are unconstrained by empirical data at very early stages. The initial specific entropy of the objects, which is dictated by details of the gas accretion process, is treated as a free parameter even though its value has a significant impact on the objectsÕ luminosity in the first few hundred million years \citep{marley2007,spiegel2012}. 

Up to now, young planet candidates have been discovered inside large gaps in circumstellar disks surrounding a few young stars \citep{kraus2012,reggiani2014,biller2014}. Located within 20 au from their hosts, they yield first luminosity estimates of young gas giant planets and suggest that near the end of their formation, giant planets have cleared disk gaps predicted by theory \citep[e.g.,][]{crida2006}. Until now, no protoplanet still embedded in the circumstellar disk from which it is forming has been confirmed. We detected a candidate protoplanet around the young star HD100546, but the single-wavelength data did not permit characterization nor was it unambiguously shown to be a true companion \citep{quanz2013a}. Recently, \citet{currie2014} recovered this object, but also this study was based on single wavelength data and could not confirm common proper motion. 

\begin{table*}
\scriptsize
\begin{center}
\caption{Summary of observations.\label{table1}}
\begin{tabular}{llllllllll}
\tableline\tableline
Date & Object & Filter & DIT\tablenotemark{a} & \# of & \# of & Airmass & Parallactic & Detector & Pixel scale\\
       &       &   &     & data & frames & & angle & window & \\
       &       &   &     & cubes & per cube & &  start / end &  & \\
\tableline\\
\multicolumn{10}{c}{High-contrast, pupil stabilized observations}\\
\tableline
2013/4/18& HD100546 & $L'$ & 0.15 s & 288 & 200 & 1.66--1.43 & -55.83$^\circ$ / +42.36$^\circ$ & 512x512 px & 27 mas/px\\

       	 	     & HD100546 & $K_s$ & 0.5 s & 90  & 100 & 1.57--1.42 &  -46.34$^\circ$ / +31.95$^\circ$ & 512x512 px & 13 mas/px\\

2013/4/19& HD100546 & $M'$ & 0.04 s & 244 & 500 & 1.61--1.43  & -50.75$^\circ$ / +24.64$^\circ$ & 256x256 px & 27 mas/px\\
		     
       	 	& HD100546    & $K_s$ & 0.5 s & 120  & 100 & 1.54--1.43 &  -41.39$^\circ$ / +32.07$^\circ$ & 512x512 px & 13 mas/px\\
\tableline
\tableline\\
\multicolumn{10}{c}{Photometric calibration observations}\\
\tableline
2013/4/18 & HD100546 & $L'$ & 0.025 s & 8 & 1200 & 1.56--1.57 &\quad\quad\quad -/- &256x256 px & 27 mas/px\\

       	 	     & HR6572 & $L'$ & 0.05 s & 6  & 600 & 1.42--1.40 &\quad\quad\quad -/-  &256x256 px & 27 mas/px\\

		     & HD100546 & $K_s$ & 0.05 s & 4 & 750 & 1.58--1.59 &\quad\quad\quad -/- &256x256 px & 13 mas/px\\

       	 	     & HR6572 & $K_s$ & 0.05 s & 4  & 750 & 1.38--1.37 & \quad\quad\quad -/-& 256x256 px & 13 mas/px\\		     	     

2013/4/19 & HD100546 & $M'$ & 0.01 s & 8 & 2000 & $\sim$1.50 & \quad\quad\quad -/-&130x130 px & 27 mas/px\\

       	 	     & HR6572 & $M'$ & 0.02 s & 16  & 1000 & 1.71--1.61 &\quad\quad\quad -/-  &130x130 px & 27 mas/px\\
\tableline
\tableline
\end{tabular}
%% Any table notes must follow the \end{tabular} command.
\tablenotetext{a}{Detector integration time, i.e., exposure time.}
\end{center}
\end{table*}

We now confirm that this object is bound to the central star and that its multi-band photometry is best explained as a newly forming gas giant planet embedded in the circumstellar disk around the young star HD100546. HD100546 has a spectral type of B9Vne \citep{levenhagen2006} and lies at a distance of $97\pm4$ pc \citep{vanleeuwen2007}. Based on photometric measurements covering wavelengths from the UV to the infrared \citet{vandenancker1997} estimated the luminosity and effective temperature of HD100546 to be $L\approx 32\;L_\sun$ and $T_{\rm eff} \approx 10500$ K from which, via comparison with stellar evolutionary models, they derived an age of $\geq$10 Myr and a mass of $2.4\pm0.1$ M$_\sun$. A slightly younger age of 5 -- 10 Myr was estimated by \citet{guimaraes2006} from high-resolution optical spectra. These authors  used the spectra to infer the effective temperature and surface gravity of HD100546 and then compared the values to stellar evolutionary tracks to get an age. As the age of the star is an important parameter in the context of (gas giant) planet formation, we will consider a range of ages between 1 -- 10 Myr throughout this paper, where the youngest age is motivated by the idea that our object might be younger than the star as it is still in the process of formation. The large ($r > 300$ au) gas and dust disk around HD100546 has been spatially resolved in scattered light as well as in thermal emission \citep{pantin2000,augereau2001,grady2001,liu2003_b,leinert2004,ardila2007,quanz2011,avenhaus2014,pineda2014,walsh2014}. The star is actively accreting material from the innermost disk regions \citep[e.g.,][]{deleuil2004,grady2005,guimaraes2006}, but there is observational evidence for a disk gap stretching from a few au out to roughly 14 au that is possibly created by an orbiting companion \citep{bouwman2003,grady2005,acke2006,benisty2010,quanz2011,avenhaus2014,brittain2013,brittain2014}. This inner companion together with the newly forming companion in the outer disk discussed in the following makes HD100546 currently the best candidate for a system, where we can directly study the formation of multiple planets and their interaction with the circumstellar disk.

\section{Observations}
We re-observed HD100546 on April 18 and 19, 2013, with the NACO instrument \citep{lenzen2003,rousset2003} installed at one of the 8.2-m Utility Telescopes of the Very Large Telescope at the Paranal Observatory of the European Southern Observatory. Data were taken in the $L'$ ($\lambda_{\rm eff}=3.770\,\mu m$) and $K_s$ ($\lambda_{\rm eff}=2.124\,\mu m$) filters in night 1, and in the $M'$ ($\lambda_{\rm eff}=4.755\,\mu m$) and again $K_s$ filters in night 2. The observations were carried out in pupil tracking mode leading to a natural rotation of the cameraÕs field of view following the changes in parallactic angle over the course of the observing sequence. Each night we switched between the different filters several times to ensure comparable field rotation in all datasets. To correct for bad detector pixels and allow for subtraction of thermal background emission, the objects were moved to different positions on the detector every 30 to 60 seconds in the $L'$ and $M'$ filter and roughly every 90 seconds in the $K_s$ filter. For the data taken in the $L'$ filter the Apodizing Phase Plate (APP) coronagraph \citep{kenworthy2010,quanz2010} was used to enhance the contrast performance of the instrument. No coronagraph was used for the $M'$ and $K_s$ filter observations. In the high-contrast datasets, in order to increase the sensitivity to faint companions, the central few pixels of the stellar Point Spread Function (PSF) were saturated, resulting in the need for additional, unsaturated datasets to determine the exact photometry of HD100546 and thus its companion. As the observing conditions were photometric in both nights, the photometric standard star HR6572 was observed as reference target. The observations are summarized in Table~\ref{table1}.

\section{Data reduction}
\subsection{Basic steps}
\emph{Pupil-stabilized data $K_s$ filter:} Individual frames were flat-fielded, dark-subtracted and bad pixel corrected (5-$\sigma$ clipping in 9 pixel box). For image alignment, a Moffat profile was fitted to a reference image and to the individual science images to determine the spatial offset between them. The individual images were then scaled up by a factor of 10, shifted by the offset and then scaled back to the original image size.

\emph{Pupil-stabilized data $L'$' filter with APP:} Individual exposures from subsequent detector positions were pairwise subtracted (to subtract background emission and dark current), and bad pixel corrected (5-$\sigma$ clipping in 9 pixel box). For image alignment, given the strong asymmetry of the APP PSF with the diffraction rings being suppressed on one side of the central star \citep{codona2006}, we cross-correlated the individual science images with a reference image to determine the spatial offset between them. The individual images were then scaled up by a factor of 10, shifted by the offset and then scaled back to twice the original image size. 

\begin{figure*}
\centering
\epsscale{1.}
\plotone{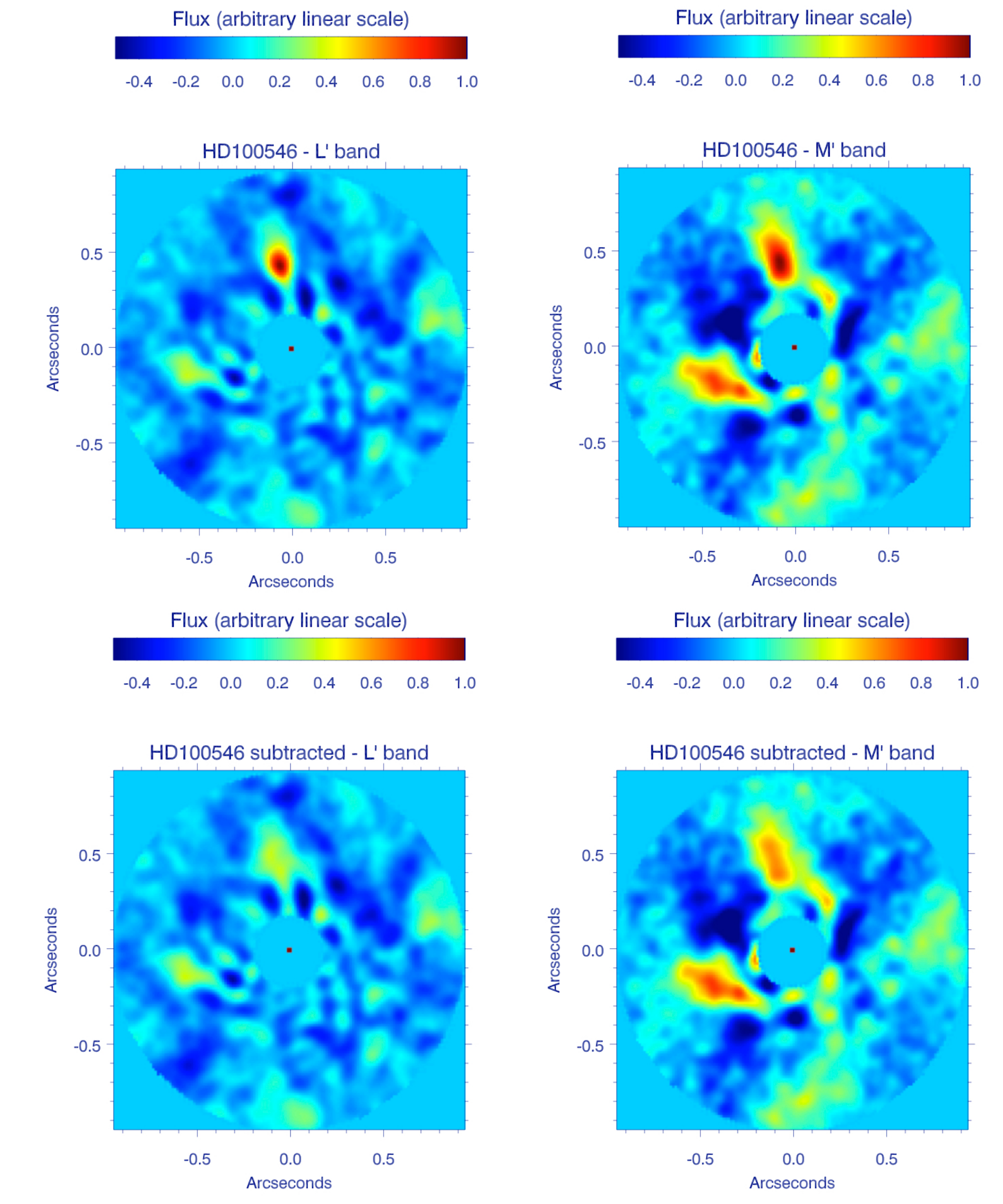}
\caption{Top row: Final PSF-subtracted images of the vicinity around HD100546 in the $L'$ filter (left panel) and $M'$ filter (right panel). The dark spot in the image center indicates the location of the central star. The innermost regions have been masked out during the data analysis as they are dominated by subtraction residuals. Bottom row: Same as above but after subtraction of the point source component. Residual resolved structures remain visible in the vicinity of the protoplanet. Additional extended emission is present to the southeast of the star mainly in the $M'$ filter (see also Section 6). North is up and East to the left in all images.\label{fig1}}
\end{figure*}

\emph{Pupil-stabilized $M'$ filter:} Individual frames were dark-subtracted and bad pixels were flagged with a bad pixel mask. Background emission was then subtracted using a novel principal component-based approach (Quanz et al. in prep): The star followed a 4-point dither pattern on the detector, where each position was roughly centered in each of the 4 detector quadrants. Hence, throughout the full stack of image cubes, for a given quadrant, the star was \emph{not} located in this quadrant $\sim$75\% of the time. These ÔstarlessÕ frames were used for the decomposition of the background emission into principal components (PCs). In order to fit the background in a given quadrant in those frames where the star was present, the innermost parts around the star were masked out and then the PCs were fitted to the remaining area. The background for a given frame was then constructed from this fit, which automatically interpolated those innermost regions containing the star that had been masked out before. For image alignment, a Moffat profile was fitted to a reference image and to the individual science images to compute the offset between them. The individual images were then scaled up by a factor of 10, shifted by the offset and then scaled back to twice the original image size.

For all filters, individual images with poor adaptive-optics correction were disregarded from any subsequent analyses. Those images were identified during the initial data reduction steps by comparing the PSF peak flux of consecutive images: in case the flux dropped by $\sim$40\% from one image to the next we flagged and visually inspected the image. In total, less than 1\% of the images suffered from bad AO performance and were rejected.

\subsection{PSF subtraction}
To reveal the existence of possible faint nearby companions, the stellar PSF is subtracted from the individual images. This was achieved using the {\sc PynPoint} data analysis package \citep{amara2012,amara2014} that is based on a PC analysis algorithm to model the light distribution of the PSF. The PCs are fitted to each individual image and the stellar PSF is subtracted. All images are then rotated to a common sky orientation, averaged and convolved with a Gaussian kernel with a full-width-half-maximum of half the size of the stellar PSF to increase the signal-to-noise (SNR) of any faint companion. The size of the stellar PSF was determined from the average of the unsaturated data sets in each filter.

\section{Observational results}
\subsection{Recovery of the protoplanet in $L'$ and $M'$}
The fully processed $L'$ and $M'$ images reveal an emission source North of the star. The source is spatially resolved in both images, but one dimensional cuts through the object's flux distribution in the radial direction reveal a point source component that dominates the flux. Hence, the object is best explained by a spatially unresolved point source component surrounded by a spatially resolved extended emission component (Figure~\ref{fig1}, top row). The total SNR of the detection was estimated following the prescription of \citet{mawet2014} and their equation 9, taking into account the peak flux of the planet and the standard deviation in a number of background pixels each representing a statistically independent resolution element at the separation from the star that is of interest. As the final {\sc PynPoint} images were convolved with a Gaussian filter, individual pixels separated enough to be statistically independent, i.e., one pixel per resolution element, were chosen as background pixels. The background pixels were selected from 2 concentric rings around the star with radii of $\sim$0.46$''$ and $\sim$0.54$''$, respectively, to have a comparable separation from the central star as the object and the surrounding extended emission component.  For the $L'$ images taken with the APP coronagraph only the high-contrast side of the APP was considered for selecting background pixels; for the $M'$ images complete circles were considered. Excluded were the immediate surroundings of the source as well as the extended emission region east of the star (see, bottom row in Figure~\ref{fig1} and section 6). This resulted in 19 and 32 background pixels for the $L'$ and $M'$ images, respectively. We then computed the SNR for a grid of different reduction parameters varying:
\begin{enumerate}
\item The number of individual images that were averaged before {\sc PynPoint} was run (5, 10 and 50 in $L'$; 10, 20 and 100 in $M'$)
\item The number of PCs used to fit the PSF (between 10 and 50 in $M'$ and between 10 and 120 in $L'$, always in steps of 10) 
\item The exact location of the reference pixels (shifting all pixels simultaneously by $\pm$1 pixel in x and y resulting in a total number of 9 sets of reference pixels)
\end{enumerate}

In all of these reductions the SNR of the object was always $\gtrsim$4 in $L'$ and $\gtrsim$3 in $M'$ and the mean SNR in $L'$ and $M'$ was $\approx$7.9 and $\approx$4.6. Irrespective of the exact location of the reference pixels, the highest SNR values were always achieved for 70 PCs and pre-averaging of 10 images in $L'$ filter (average SNR$\approx$11.4)\footnote{We note that the SNR  of the $L'$ detection in the discovery paper \citep[SNR$\approx$15;][]{quanz2013a} was overstimated as we did not carry out a rigorous SNR assessment as we did here (incl. varying in the reduction parameters and background pixels, correcting for any offset in the mean level of the background pixels, etc.)} and for 10 PCs and pre-averaging over 100 images in the $M'$ filter (average SNR$\approx$6.6). The resulting images based on these parameter were used for all subsequent analyses and are shown in Figure~\ref{fig1}.

We note that given the small number of background pixels it is not possible to robustly constrain their underlying distribution without making some ad-hoc assumption of their general functional form. In consequence, without assuming a functional form for the underlying distribution, we cannot assign a confidence level to our detection. However, given that the object was already detected previously in independent datasets \citep{quanz2013a,currie2014} and is now re-detected in two different filters makes us believe that a statistical outlier or instrumental artifact is extremely unlikely.

\subsection{Contrast and astrometry of the protoplanet in $L'$ and $M'$}
The PSF subtraction step affects the exact location and brightness of any companion. To estimate the protoplanet's contrast and position, artificial negative objects (covering a grid of varying brightness (in steps of 0.1 mag) and location (in steps of 0.25 pixels in x and y, respectively)) were inserted in the individual input images and {\sc PynPoint} was re-run. 
An unsaturated PSF-core from the photometric datasets was used as template to create the fake sources. To determine their flux levels in the saturated datasets differences in airmass during the observations and the difference in exposure time between the saturated and unsaturated datasets were taken into account. To subtract the stellar PSF we used the same number of PCs as in the final images defined above. However, a new set of PCs was constructed for each new dataset containing fake native planets. Based on the previous results from the initial discovery paper \citep{quanz2013a} and from the final images shown in Figure~\ref{fig1}, the baseline assumption was that the emission from the companion consists of a point source component surrounded by spatially resolved, extended emission component. 
To determine the contrast and location of the point source component the goal was to subtract only so much of the emission so that the remaining flux is comparable to the flux level in the immediate vicinity of the object (i.e., the extended emission component) without creating strong structural asymmetries in the final image. To evaluate the impact of removing flux from the point source we quantify the level of curvature at the objectÕs position using the determinant of the Hessian matrix of the image at that point. In an analogous way to second derivatives in 1D, excess flux from a compact source would lead to a positive determinant while an over subtraction would lead to a hole and thus a negative determinant. This allowed us to determine the best-fit subtraction, which was confirmed through visual inspection: the residual images, once the flux from the point source had been removed in this way, looked consistent with a smooth broader background emission region (see, Figure~\ref{fig1} bottom row). For further discussion of image feature detection using Hessian matrix methods and others we refer the reader to, e.g., \citet{lindeberg1998} and \citet{bay2008}.

For the $L'$ images this resulted in a contrast of $\Delta L' = 9.4 \pm 0.1$ mag between HD100546 and the compact emission component. The accuracy of the location of the compact emission component was $\pm$0.5 pixels in the {\sc PynPoint} images, which translates into $\pm$0.25 pixels in the original image size (see Section 3.1 above). This error does not include systematic uncertainties if we had chosen a different final image (i.e., with a different number of PC used to subtract the stellar PSF, see above). Fake sources that are brighter/fainter or more offset than the values quoted above left a measurable level of curvature and clearly visible brightness asymmetries in the final subtracted images. The same analysis carried out for the $M'$ data yielded a contrast of $\Delta MÕ= 9.2 \pm 0.15$ mag between HD100546 and the compact emission component and the same accuracy in the object's location. 

It is worth noting that after subtraction of the best-fit negative point source the remaining extended emission is not only elongated in the radial direction in the final images, but now also left and right of the point source component persisting emission structures appear (see, Figure~\ref{fig1} bottom row). This can be understood if one keeps in mind that normally any reduction algorithm for pupil stabilized images creates negative ÔholesÕ left and right of a point source as it tries to cancel it out. In the final {\sc PynPoint} images no negative holes are present, but the source has an unusually elongated shape, which we interpret as {\sc PynPoint} canceling out flux left and right of the point source component. Part of this flux reappears after the point source component is subtracted. Without extensive modeling the precise shape and brightness distribution of the extended emission component cannot be determined, however, and we basically assume that it is ÔflatÕ for the analysis described here. 

The final astrometry of the object was derived using the location of the best-fit negative fake planet and the location of the central star (with an uncertainty of $\pm$0.2 pixels in x and y) and we found $(0.457\pm0.014)''$ and $(0.472\pm0.014)''$ for the separation and $(8.4\pm1.4)^\circ$ and $(9.2\pm1.4)^\circ$ for the position angle in the $L'$ and $M'$ filter, respectively. These values are consistent with those found in our discovery paper \citep{quanz2013a}. The errors are calculated from error propagation in $r$ (the separation derived from $\Delta x$ and $\Delta y$ between planet and star) and in $\Delta x / \Delta y$ (which is used to compute the position angle via $PA = \tan^{-1}(\Delta x/ \Delta y))$. A systematic error in the position angle from an unknown true north orientation of the camera field-of-view is not included, but is estimated to be $<1^\circ$. Assuming an inclination and position angle of the circumstellar disk of 42$^\circ$ and 145$^\circ$, respectively \citep{pineda2014}, the average de-projected physical separation of the compact object is $(53\pm2)$ au. The error assumes the same uncertainties in $r$ and $PA$ as for the individual measurements of these parameters in $L'$ and $M'$ band and denotes the results for the most extreme combinations of $r$ and $PA$ allowed in the given range. Uncertainties in the disk inclination and disk position angle are not included.

\begin{figure*}
\plotone{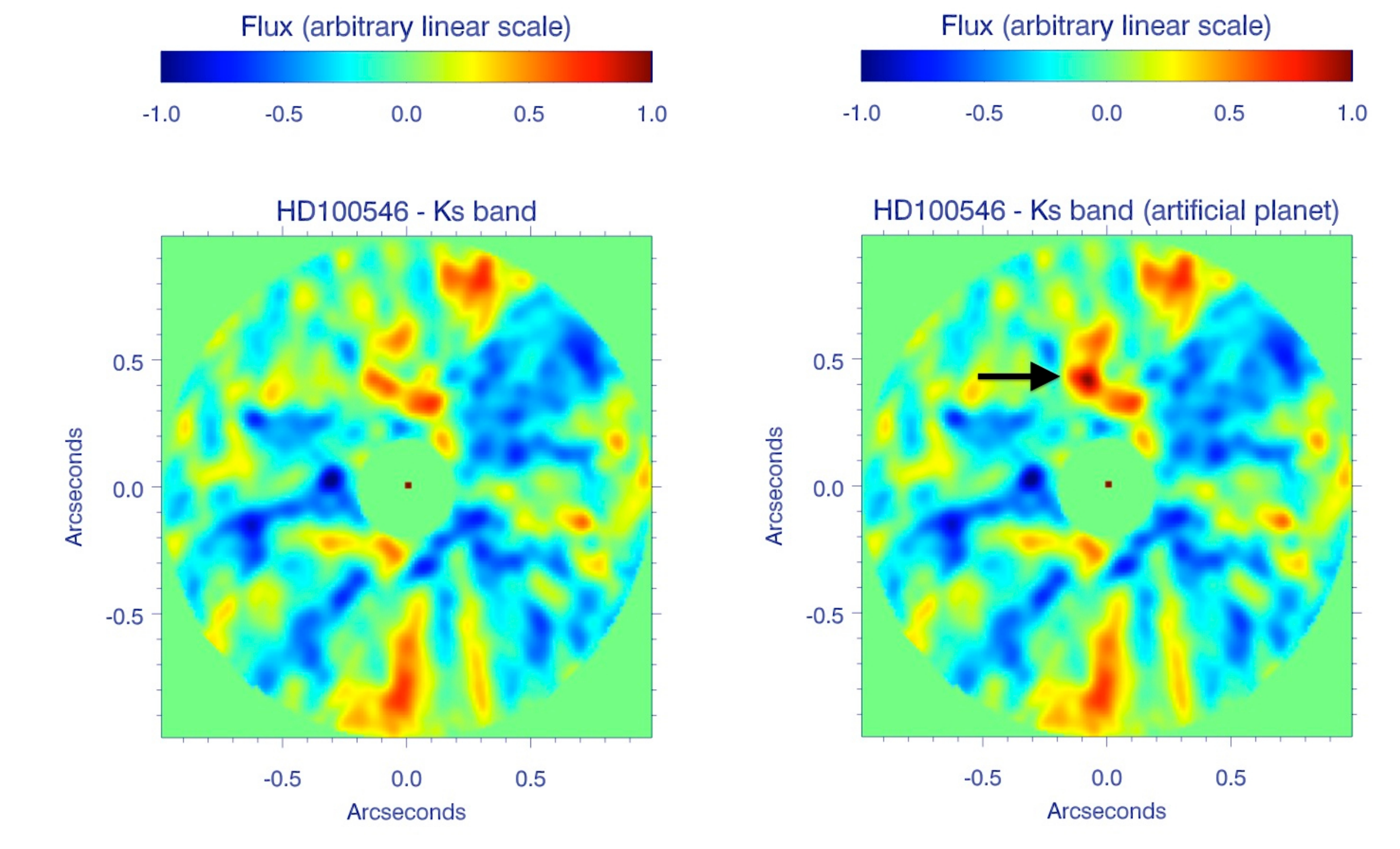}
\caption{Derivation of the upper flux limit in the $K_s$ filter. Left: Final $K_s$ image reduced with 10 principal components in {\sc PynPoint}. No significant point source is detected. Right: Same image with an artificial planet with $\Delta K_s = 9.6$ mag inserted in the input data at the expected location of the companion. The object is detected with an SNR $\sim$ 3 (black arrow). North is up and East to the left. \label{fig2}}
\end{figure*}

\subsection{Non-detection of the protoplanet in $K_s$}
While the protoplanet is clearly detected in the $L'$ and $M'$ filter, the object is not detected in any of the two $K_s$ datasets that were obtained. To estimate an upper limit of the objectÕs $K_s$ brightness we inserted fake planets with varying brightness (in steps of 0.1 mag) at the expected location derived from the $L'$ and $M'$ data and re-ran {\sc PynPoint}. For this we used the $K_s$ dataset from night 1 as overall the observing conditions and the AO performance were better compared to night 2. To compensate for a lower Strehl ratio and higher variability in the AO correction in the $K_s$ filter compared to $L'$ or $M'$, we selected the best 50\% of the image frames from night 1. For this we subtracted the mean $K_s$ image from night 1 from all individual images and computed the residual noise in the resulting image by determining the standard deviation of all pixels. We then sorted all images by their noise level and took the best 50\%, i.e., those with the lowest noise, as our input images for further analyses.
To estimate the upper limit for the $K_s$ brightness we applied the same approach as done for the $L'$ and $M'$ detections \citep[see, section 4.1; cf.][]{mawet2014}. We estimated the SNR of fake companions for different numbers of PCs and 2 different sets of input data: one, where each of the selected images was used, and one, where 10 consecutive images were stacked before {\sc PynPoint} was run (similar to the analysis in $L'$ and $M'$). As background pixels we used 24 statistically independent pixels from a concentric ring around the central star with a radius corresponding to the separation between star and protoplanet. We found that for a contrast of $\Delta K_s = 9.6$ mag a fake companion would have been detected with an average SNR of $\approx 3.7$ in both sets of input data for reductions with 5, 10, 20, and 40 PCs (see, Figure~\ref{fig2}). Given that we know from the $L'$ and $M'$ datasets that there is a source at this location we use this contrast as our 3-$\sigma$ upper limit. 

We note that already \citet{boccaletti2013} did not detect the protoplanet in the $K_s$ filter in archival data from Gemini/NICI. The detection limits we achieve here are somewhat deeper, further emphasizing the very red colors of the object (see below).

\subsection{Photometry of HD100546 and the protoplanet}
We observed HR6572 (an A0V star) as photometric standard star in both nights as the observing conditions were photometric. Unfortunately, HR6572 has no listed $M'$ magnitude and hence we assumed that $M' = L'$. For the photometry, each data cube (see, Table~\ref{table1}) was reduced and analyzed individually including bad pixel cleaning, background subtraction, image alignment and averaging (see, Section 3.1). The stellar flux was measured in the final image in an aperture with 3 pixels radius in the $K_s$ and $L'$ filter, and in an aperture with 4 pixels radius in the $M'$ filter. The average flux of all final images was taken as the final flux and the standard deviation of the flux of all final images divided by the square root of the number of final images was taken as error on the flux measurement (i.e., the standard deviation of the mean). The final apparent magnitudes for HD100546 were obtained by comparing its final fluxes in the different filters to those of HR6572 (normalized to the same integration time) and using the cataloged magnitudes of HR6572 as reference points. To correct for the difference in airmass between HD100546 and the HR6572, we used the Paranal extinction values listed on the ESO/NACO webpage ($K_s$: 0.07 mag, $L'$: 0.08 mag, $M'$: 0.15$\pm$0.05 mag\footnote{For the $M'$ filter ESO provides a range for the extinction value of 0.1--0.2 mag. We chose to use the midpoint of this range and included an error term in subsequent analysis to reflect the uncertainty.} ). The final photometric errors (in magnitudes) for HD100546 were computed via
\begin{eqnarray*}
\sigma^{M'}_{\rm total, HD100546} = \Big((\sigma^{M'}_{\rm obs,HD100546})^2 + (\sigma^{M'}_{\rm obs,HR6572})^2 \\+ (\sigma^{M'}_{\rm cat,HR6572})^2 + (\sigma^{M'}_{\rm airmass})^2\Big)^{1/2}
\end{eqnarray*}

with 
\begin{description}
\item {$\sigma^{M'}_{\rm obs,HD100546}$:} observed standard deviation of mean flux of HD100546
\item {$\sigma^{M'}_{\rm obs,HR6572}$:} observed standard deviation of mean flux of HR6572
\item {$\sigma^{M'}_{\rm cat,HR6572}$:} photometric error of HR6572 in catalog
\item {$\sigma^{M'}_{\rm airmass}$:} error in airmass correction.
\end{description}

For $L'$ and $K_s$ the magnitude errors were computed accordingly, with the only difference being that no error in the airmass correction was included. Using the contrast measurements from the previous section and the apparent magnitudes for HD100546, we derived the apparent magnitudes of the compact emission component, i.e., the protoplanet. For this, we also investigated how color correction terms due to the red color of the object and the different filter systems in NACO and in the reference star catalog might affect the results. We found this effect to be $<$2\%, which is smaller than the final uncertainties in the apparent magnitudes of the object. We summarize the photometric results in Table~\ref{table2}. The errors for HR6572 comprise the second and third term from the right hand side of the equation above.

\begin{table}
\begin{center}
\caption{Observed and derived apparent magnitudes.\label{table2}}
\begin{tabular}{llll}
\tableline\tableline
Object & $K_s$ [mag] & $L'$ [mag] & $M'$ [mag] \\
\tableline
HR6572 & 5.738$\pm$0.016 & 5.726$\pm$0.009 &	5.73\tablenotemark{a}$\pm$0.014 \\
HD100546 & 5.83$\pm$0.06 & 4.52$\pm$0.02 & 4.13$\pm$0.05 \\
HD100546 b &	$>$15.43\tablenotemark{b}$\pm$0.06 & 13.92$\pm$0.10 & 13.33$\pm$0.16 \\ 
		     & $>$13.59\tablenotemark{b,c}$\pm$0.06&	12.75\tablenotemark{c}$\pm$0.10 & 12.33\tablenotemark{c}$\pm$0.16\\
\tableline
\end{tabular}
\tablenotetext{a}{HR6572 (an A0V star) has no listed $M'$ magnitude in the standard star catalog, hence we assume $M' \approx L' (\approx K_s)$, but increase the associated error.}
\tablenotetext{b}{3$\sigma$ detection limit; the error reflects the uncertainty in the photometry of HD100546.}
\tablenotetext{c}{Values after correcting for dust extinction effects (section 5.3.2.)}
\end{center}
\end{table}

\section{Analysis}
Throughout the paper we already referred to the detected point source component as 'protoplanet'. In the following we show that its observed properties are indeed best explained with a young, forming planet.

\subsection{Rejection of fore-/background objects}
\begin{figure}
\plotone{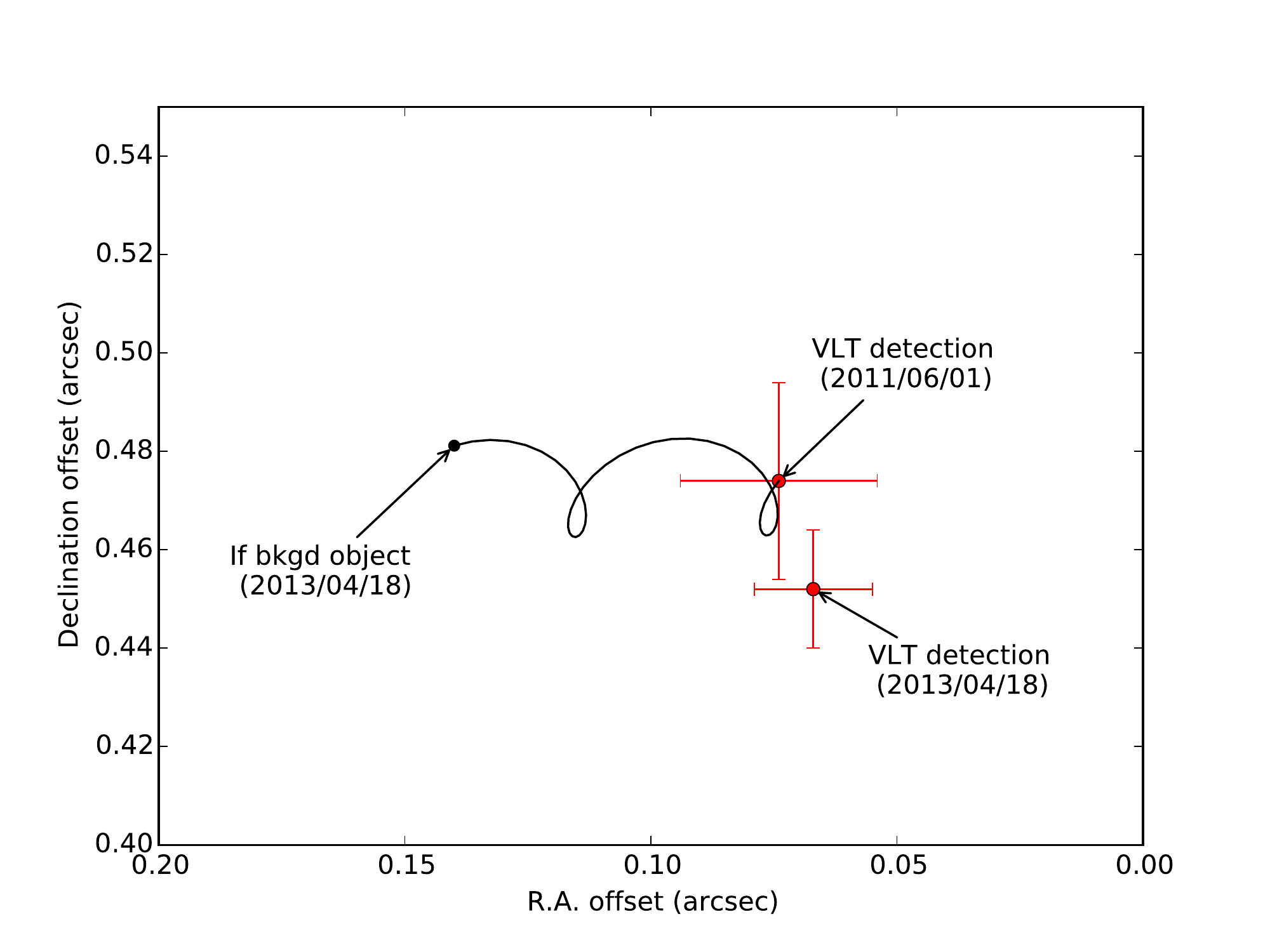}
\caption{Proper motion analysis based on the observed astrometry of the protoplanet in the two epochs. The x and y axes show the offset with respect to the central star. The red crosses denote the location of the protoplanet relative to the star in the two epochs. The black line shows the expected motion of a background source between the two epochs based on the parallactic motion and proper motion of HD100546.\label{fig3}}
\end{figure}

To check for common proper motion between the protoplanet and HD100546 the $L'$ data presented here we taken with exactly the same observational setup as the data presented in the initial discovery paper \citep{quanz2013a}. A proper motion analysis based on the measured astrometry of the protoplanet and the parallactic motion and proper motion of HD100546 between the two epoch shows that the object is inconsistent with a stationary background source (see, Figure~\ref{fig3}). The analysis assumes a proper motion of HD100546 in right ascension and declination of $-38.93$ mas/yr and $0.29$ mas/yr, respectively \citep{vanleeuwen2007}. The astrometric accuracy of the data presented here is discussed above. A re-analysis of the data from epoch 1 led to an uncertainty in the protoplanetÕs location of $\sim$0.75 pixels in x and y (incl. uncertainties in the location of the central star). We made sure that both datasets, from epoch 1 and 2, were aligned to the same reference image to minimize systematic offsets in determining the location of the protoplanet. As the data presented here have higher SNR than the epoch 1 data, they provide a better precision in the protoplanetÕs location.

In addition to this, the combination of apparent $L'$ and $M'$ band brightness and color are inconsistent with those of any stellar foreground or background object. Only some very cool brown dwarfs with spectral types of T6 and later show similar properties, but, without exception, these objects are located in the immediate neighborhood from the Sun with distances typically $<$10 pc and hence very large proper motions \citep{golimowski2004,faherty2009}. Given that the location of our object has not changed compared to data from 2011 we can exclude that it is a nearby object. Rather, the red color of the compact component combined with the extended emission component suggests that the object is associated with the circumstellar disk of HD100546. 

\subsection{Rejection of scattered light from the circumstellar disk}
As HD100546 is surrounded by a large, flared circumstellar disk that has been detected in scattered light at multiple wavelengths, the detected emission (point source + extended component) could, in principle, be starlight reflected from the disk's dusty surface layer. However, using polarized light as tracer for scattered light, this seems rather unlikely. No local brightness maximum is seen in high-contrast polarized light images of HD100546 at the location of the object, neither in the NIR at $H$ or $K_s$ nor in $L'$  \citep{quanz2011,avenhaus2014}.

\subsection{Effective temperature and emitting area}
\subsubsection{Without dust extinction effects}
Rejecting scattered light as origin for the observed flux, leaves thermal emission coming from (within) the circumstellar disk as a possible source. As derived above the de-projected physical separation of the compact object is $(53\pm2)$ au. At this separation from the central star, radiative transfer models predict a temperature of $\sim$50 K in the mid-plane of the HD100546 circumstellar disk \citep{mulders2011}, which is inconsistent with the observed $L'-M'$ color. A local extra source of energy is required. 

Assuming a distance of  ($97\pm4$) pc, we used blackbody emission to estimate the effective temperature and emitting area of the detected point source component. We computed blackbody fluxes for a 2D grid of temperatures $T_{eff}$ (from 500 to 3000 K in steps of 5 K) and radii $R$ (from 1 to 25 R$_{\rm Jupiter}$ in steps of 0.1 R$_{\rm Jupiter}$) and convolved them with the NACO filter transmission curves. We then computed a $\chi^2$-grid (each cell corresponding to a certain $T_{eff}-R$ combination) by fitting the blackbody fluxes  to the observed fluxes\footnote{We assumed the following zero points (in erg/cm$^2$/s/\AA): $4.501\cdot10^{-11}$, $5.151\cdot10^{-12}$ and $2.117\cdot10^{-12}$ for $K_s$, $L'$, and $M'$, respectively (see, http://svo2.cab.inta-csic.es/theory/fps3/)} and converted this grid into a likelihood grid where the likelihood in each cell is $p \propto \exp(-\chi^2/2)$. In these fits the non-detection in the $K_s$ filter was explicitly taken into account by measuring the average flux at the expected location of the planet in the 8 final $K_s$ images used in the SNR analysis described in section 4.3. This flux was used as 'observed' flux in the $K_s$ filter. In all the blackbody fits, uncertainties in the distance estimate and photometry of HD100546 as well as in the photometry of the protoplanet (in case of the $K_s$ filter the upper limit) were taken into account.

\begin{figure*}
\centering
\epsscale{.9}
\plotone{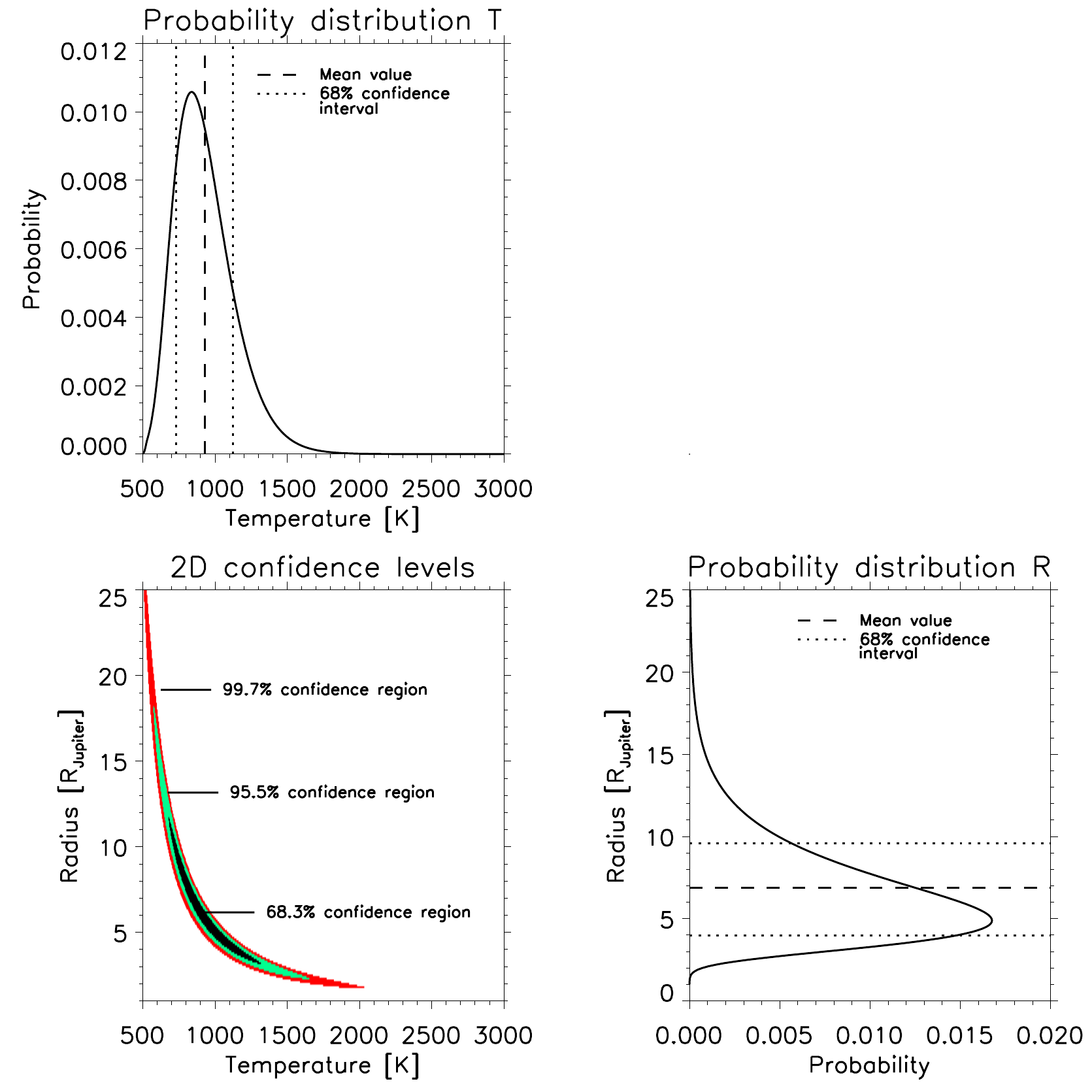}
\caption{Results from the $\chi^2$-fits of blackbodies with varying temperature and size to our $K_s$, $L'$ and $M'$ photometry. 2D confidence levels are shown in the bottom-left panel with the black, green and red colored regions denoting specific confidence regions. The probability distributions for $T_{eff}$ and $R$ are shown in the top and bottom-right panels, respectively, with the dashed lines indicating the mean value and the dotted-lines enclosing the $\sim$68\% confidence interval.
\label{fig4}}
\end{figure*}

The probability distributions for $T_{eff}$ and $R$ were computed by marginalizing over the other parameter in the likelihood grid and normalizing the resulting distribution. These were then used to compute the expectation values and confidence levels for $T_{eff}$ and $R$. This exercise yielded $T_{eff}=932^{+193}_{-202}$ K and $R=6.9^{+2.7}_{-2.9}$ R$_{\rm Jupiter}$ for the effective temperature and radius of the emitting area, respectively (the error bars define the $\sim$68\% confidence interval).

We also normalized the full likelihood grid to identify those combinations in the $T_{eff}-R$ space that corresponded to certain confidence regions (Figure~\ref{fig4}). The best-fit total luminosity of the compact component is $L=(2.3^{+0.6}_{-0.4})\cdot10^{-4} L_\sun$ and is based on the best $\chi^2$ value from the simultaneous fit of $T_{eff}$ and $R$. The bounds are the minimum and maximum luminosities found in the 1$\sigma$ contour of the combined  $\chi^2$-fit. The contour levels were derived from sorting all entries in the likelihood grid and determining those values of the likelihood, where the cumulative sum up to these values correspond to certain confidence levels (e.g., 1$\sigma$ corresponds to $\sim$68\% confidence). 

\begin{figure}
\centering
\epsscale{1.2}
\plotone{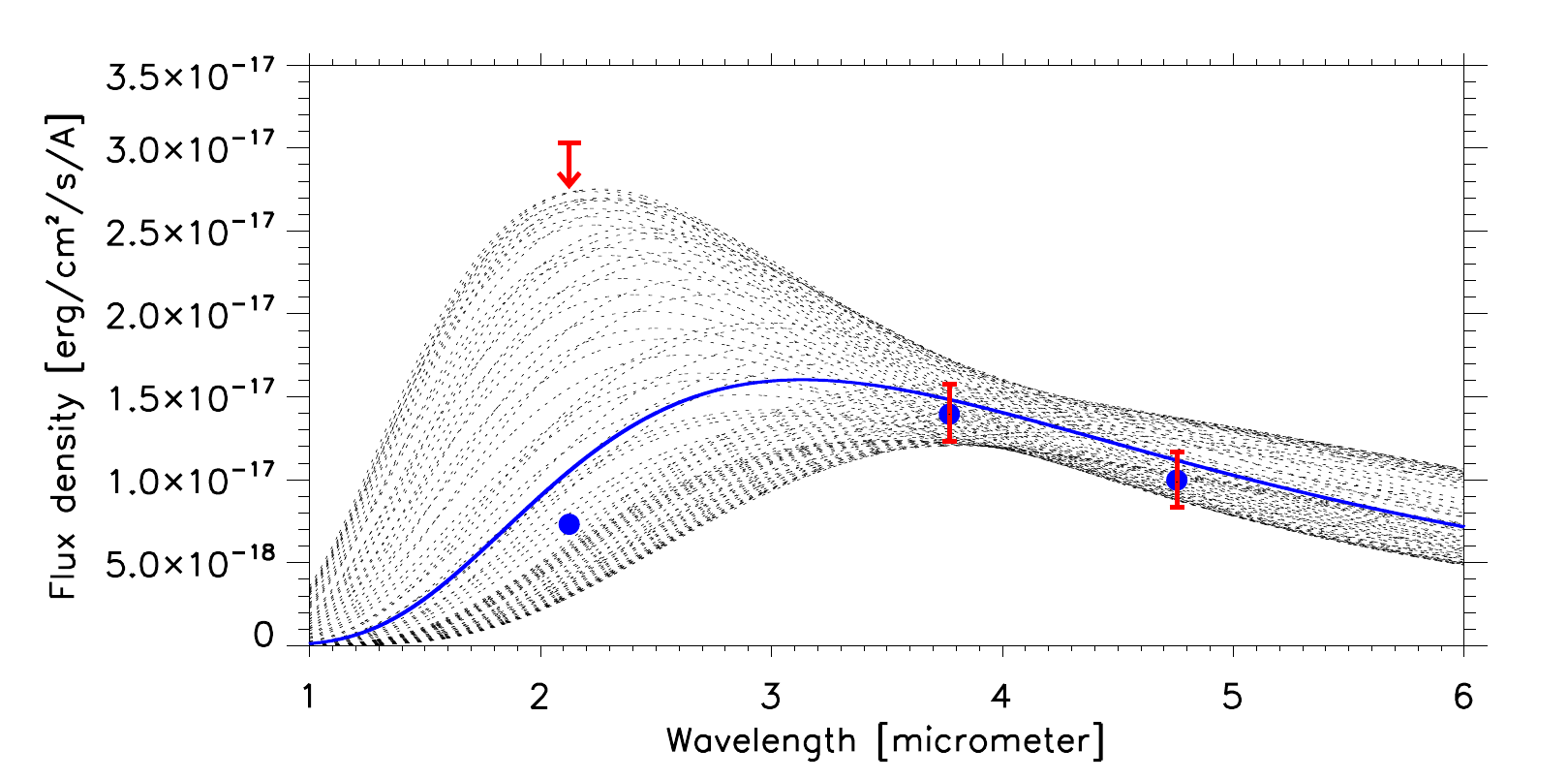}
\caption{Observed flux densities (with 1$\sigma$ error bars) and upper limit (all red data points) over-plotted on blackbody emission curves. The blue line corresponds to the best $\chi^2$-fit for effective temperature and emitting area. The blue points result from convolving the blue line with the transmission curves of the $K_s$, $L'$ and $M'$ filters. The black-dotted lines are a set of representative blackbodies from within the 1$\sigma$ region of the $\chi^2$-fit.\label{fig5}}
\end{figure}

In Figure~\ref{fig5} we show the spectral energy distribution of the best-fit blackbody, the observed fluxes as well as a set of representative blackbody curves from within the 1$\sigma$ region.

\subsubsection{Including dust extinction effects}
In case the object is embedded in the mid-plane of the HD100546 circumstellar disk, the dust between the mid-plane and the disk surface layer reduces the observed brightness of the source due to wavelength dependent scattering and absorption efficiencies of dust grains. We do not have spatially resolved information about the 3D disk structure and the local dust grain properties, but we used a radiative transfer disk model for HD100546 \citep{mulders2011,pineda2014} to estimate the optical depth of the disk at the location of the protoplanet. The values were 2.52, 1.60 and 1.37 for the $K_s$, $L'$ and $M'$ filters, respectively, and referred to the optical depth through the full face-on disk at $\sim$50 au. Using these estimates, the observed disk inclination and assuming that the object sits in the disk mid-plane we derived extinction corrected magnitudes for the protoplanet (see, Table~\ref{table2}). With the extinction corrected values for the magnitudes, we re-ran the blackbody fits described above and found an effective temperature of $T_{eff}=1242^{+353}_{-357}$ K and a radius of the emitting area of $R=7.3\pm3.2$ R$_{\rm Jupiter}$. These extinction-corrected estimates are consistent within the error bars with the uncorrected values derived in the previous section. 

We emphasize that the adopted values for the optical depth are entirely based on a circumstellar disk model and with the data in hand we cannot constrain any possible extinction effects empirically. Furthermore, simulations suggest that the formation process of a gas giant planet and its interaction with the surrounding circumstellar disk is a three-dimensional process \citep[e.g.,][]{gressel2013}, indicating that for more realistic extinction estimates other effects need to be taken into account as well. However, as the radiative transfer model is able to reproduce a large number of observational constraints, we consider this analysis to be an interesting comparison with the default results based on the observed fluxes.

\subsection{Comparison with evolutionary and atmospheric models for young planets}
Given the available data, the observed morphology and derived parameters are best explained with a young, potentially still forming, gas giant planet \citep[cf.][]{quanz2013a,currie2014}. Further evidence for a source orbiting HD100546 around 50 au comes from recent observations with the Atacama Large Millimeter Array (ALMA) showing that the distribution of mm-sized dust grains in the circumstellar disk mid-plane hints towards dynamical interactions with a young protoplanet \citep{pineda2014,walsh2014}.

\begin{figure}
\centering
\epsscale{0.9}
\plotone{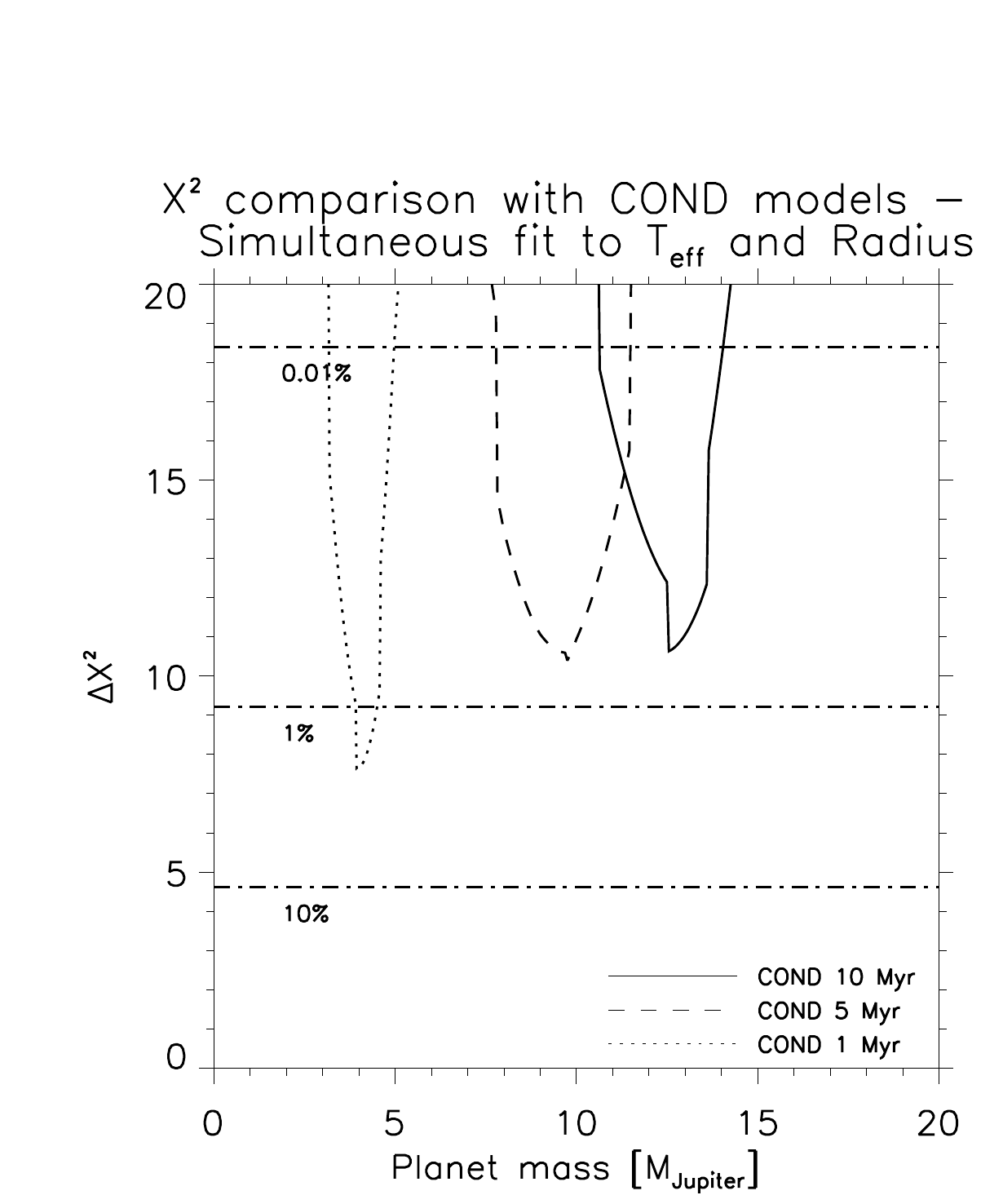}
\caption{Quantitative $\chi^2$ comparison of results from our blackbody fits with temperature-size predictions from theoretical models \citep{baraffe2003} for varying masses and ages of young gas giant planets. The dash-dotted line shows different confidence levels.\label{fig6}}
\end{figure}

\begin{figure}
\centering
\epsscale{0.9}
\plotone{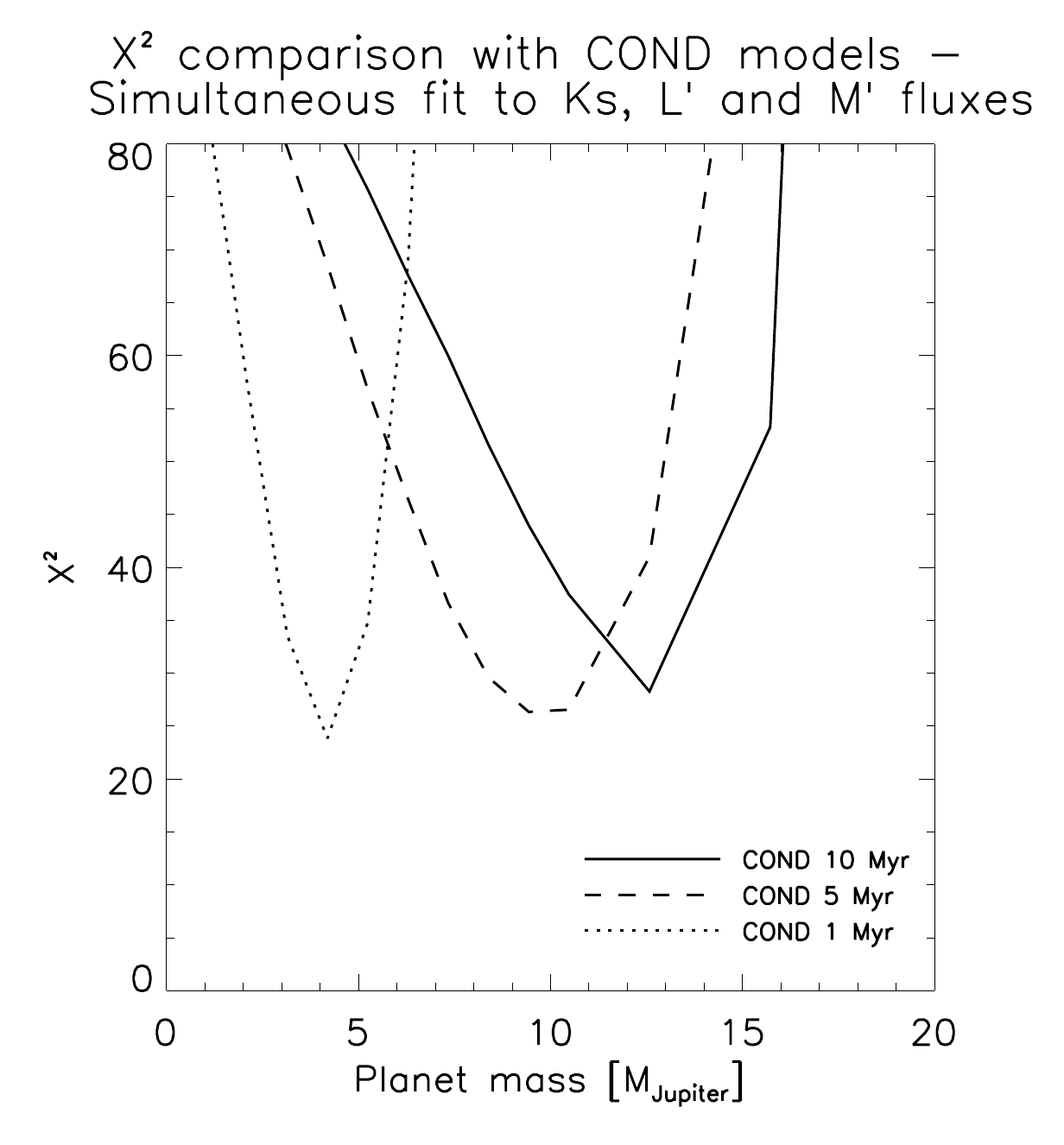}
\caption{Results from a $\chi^2$-fit of the predicted $K_s$, $L'$ and $M'$ fluxes of young gas giant planets with varying masses and ages  \citep{baraffe2003}  to our observed fluxes.\label{fig7}}
\end{figure}

As our object is presumably the youngest exoplanet discovered so far, it allows for a direct comparison with model predictions for the earliest stages of gas giant planet formation. Models with high values for the initial entropy \citep["hot-start`` models; e.g.,][]{marley2007,baraffe2003} predict combinations of radius and effective temperature that agree with our derived parameters  at a confidence level of a few percent or even less in the first 10 Myr of their evolution (see, Figure~\ref{fig6}). The best fit is found for an object with $\sim$5 M$_{\rm Jupiter}$ at an age of 1 Myr. For older ages, the best fits predict higher masses. In general, the models predict smaller radii in the relevant temperature range \citep{baraffe2003,fortney2008}. As we will discuss in section 6, one way to explain the large effective radius of the protoplanet is to assume the existence of a spatially unresolved circumplanetary disk that contributes to the detected flux. All radius-temperature fits formally improve if the extinction effects discussed above are considered: The best-fit model based on the extinction-corrected values is found for a 1 Myr-old $\sim$10 M$_{\rm Jupiter}$ object with a confidence level of $\sim$10\% (cf. Figure~\ref{fig6}).

\begin{figure*}
\centering
\epsscale{1.1}
\plotone{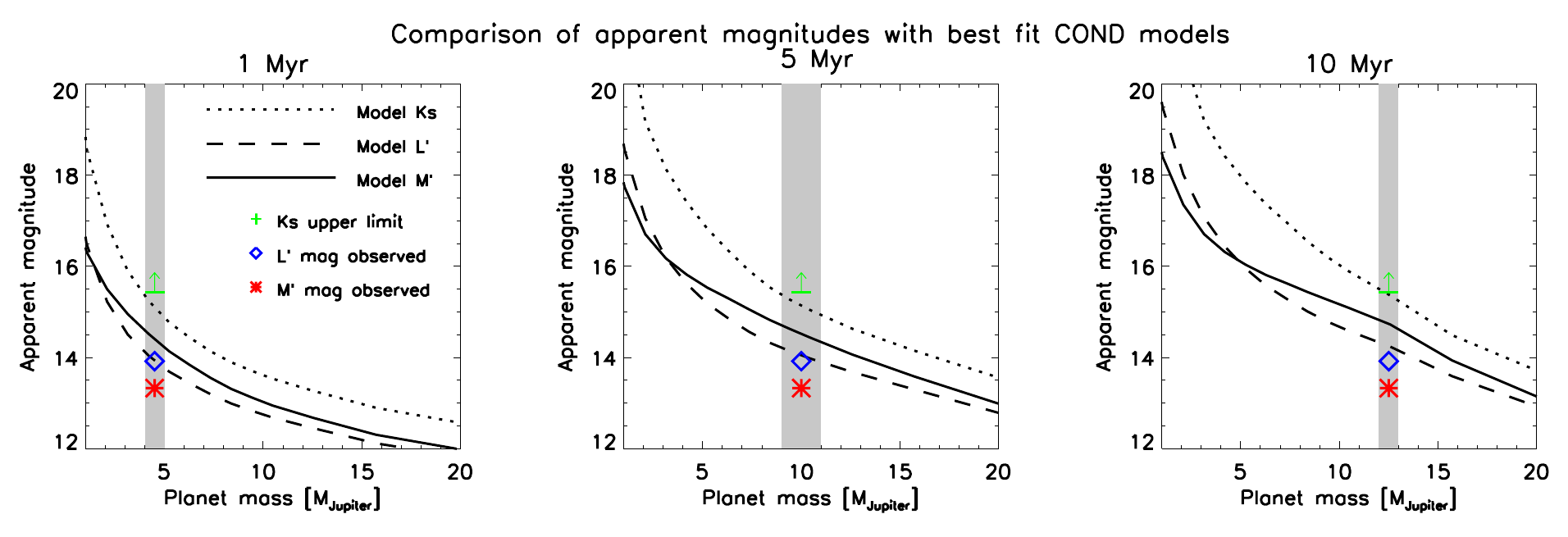}
\caption{Direct comparison of $K_s$, $L'$ and $M'$ magnitudes as predicted by the COND models \citep{baraffe2003} with the observed magnitudes of the protoplanet for ages of 1, 5 and 10 Myr (from left to right). The grey-shaded region in each panel, where the comparison is shown, is defined by the best-fit mass range from the $\chi^2$-fits presented in Figures~\ref{fig6} and~\ref{fig7}. The line and symbol legend shown in the left panel is valid for all panels. \label{fig8}}
\end{figure*}

Going one step further, we also fitted the predicted $K_s$, $L'$ and $M'$ fluxes from the COND models for a range of masses and ages of 1, 5 and 10 Myr. It is important to remember that the predicted fluxes result from the evolutionary models combined with additional predictions from the atmospheric models. These fits yield $\chi^2$ values $\gtrsim$25 (Figure~\ref{fig7}), which is significantly worse than fitting radius and temperature from the evolutionary models alone. However, it is interesting to point out that for a given age the best-fit mass range is basically identical to the best-fit mass range found in Figure~\ref{fig6}. In Figure~\ref{fig8} we illustrate the main differences between the model predictions and the observed fluxes for those regions of the mass--age parameter space that formally provide the best fit in Figures~\ref{fig6} and~\ref{fig7}. It shows that, irrespective of the assumed age, the models predict $K_s$ magitudes that are $\gtrsim$3$\sigma$ discrepant and hence should have led to a detection. Furthermore, in all cases the predicted $L'$ magnitudes are brighter than the $M'$ magnitudes, which is not observed.

Turning to models with low values for the initial entropy ("cold-star`` models), typically the predicted luminosities agree with our derived value only during a short phase ($\leq$0.1 Myr) at the beginning and at the end of the gas runaway accretion \citep{marley2007}. This would mean we have caught the object exactly at the right time, which seems unlikely. More recent work suggests that if the solid core of the planet consists of several tens of Earth masses, the resulting luminosity might be comparable to what we found here even a few million years after formation \citep{mordasini2013}. However, the predicted radii of the objects are again much smaller than what we derived above.

\section{Discussion}
\subsection{The protoplanet HD100546 b}
One of the most interesting questions is whether we can put some constraints on the mass of the protoplanet. While all "hot-start`` models predict masses of at least 5 M$_{\rm Jupiter}$, additional observational results question the presence of a high-mass protoplanet. A massive planet orbiting within a circumstellar disk opens up a gap with a width of several times the planetÕs Hill radius ($>$15 au in this case) within a few hundred orbits \citep{lin1993}. However, polarized light images of the disk show no direct evidence for a disk gap \citep{avenhaus2014}. As the orbital time of the object is only $\sim$250 years, this would imply that the object is not very massive, quite young or both. Alternatively, the local properties of the gas disk might suppress gap formation on the disk surface (e.g., due to local turbulent viscosity) or the combination of disk flaring and inclination complicates the detection of a disk gap in scattered light. Higher spatial resolution observations in the future, either in scattered light or with ALMA, can help us to search for clear gap signatures at the object's location, which could then be used to put some constraints on the object's mass.

Concerning the discrepancies between the observations and the model predictions, the presence of a circumplanetary disk, as expected from hydro-dynamical simulations of forming gas giant planets, helps to circumvent some of the problems. Such a disk extends out to roughly 40-50\% of the planetÕs Hill sphere \citep[e.g.,][]{martin2011,gressel2013} and would add an additional emission component to the system. Assuming a 2 M$_{\rm Jupiter}$ protoplanet its circumplanetary disk would be $\sim$1.4 au in radius, which would not be spatially resolved in our images and would hence be part of the compact emission component. In this scenario, our derived radius, but of course also the derived effective temperature, are then the superposition of the emission coming from the protoplanet and its disk. Such a two-component model can fit the data for several reasonable combinations of sizes and effective temperatures and data points at additional wavelengths -- preferably at longer wavelengths \citep{zhu2015,eisner2015} -- are needed to help break existing degeneracies. In principle it is possible to artificially fix the effective temperature, mass and radius of the young planet, e.g., by selecting one of the COND models, and then use the observed data to constrain the properties of the circumplanetary disk. However, the COND models (just like any other evolutionary model) are highly uncertain and unconstrained by empirical data at very early ages and furthermore, given the results in section 5.4., we have no good metric to decide, which planet model, in terms of age and mass, we should pick.

The discovery of a young gas giant planet still embedded in the circumstellar disk at $\sim$50 au from its star suggests that these objects can form at large separations. This is in particular interesting for other, slightly older, exoplanet systems, where massive gas giant planets have been detected at comparable orbital separations \citep[HR8799 bc, HD95086 b, GJ504 b;][]{marois2008,rameau2013a,rameau2013b, kuzuhara2013}. Also some of these objects may have formed close to their current location and additional mechanism, such as significant outward migration or dynamical scattering, may not be required to explain their orbits. However, both the classical core accretion model and the gravitational instability model cannot easily explain the data presented here. In the classical core accretion model, the time required to build up a rocky core of several Earth masses in situ at the given distance from the star exceeds by far the age of the system \citep[e.g.,][]{kennedy2008}. In the gravitational instability model, the disk has to be massive enough to locally fragment. The remaining mass available in the HD100546 circumstellar disk \citep[$\sim$10 M$_{\rm Jupiter}$, ~0.4\% of the stellar mass;][]{panic2010} is certainly not sufficient for fragmentation to occur, which - to first order - sets in for masses of $\sim$10\% of the stellar mass \citep[e.g.,][]{lodato2008}. So even if the disk mass was a factor of a few higher prior to the formation of the protoplanet, disk fragmentation seems unlikely. Recently it was suggested that the accretion of cm-sized pebbles that are loosely coupled to the gas in the circumstellar disk could significantly speed up the timescales for growing a rocky core even at large separation from the central star \citep{lambrechts2012}. Qualitatively, such a model seems to be able to explain the formation of a gas giant planet at $\sim$50 au, and HD100546 might be an ideal laboratory to study alternative formation processes for gas giant planets.

\begin{figure}
\centering
\epsscale{1.}
\plotone{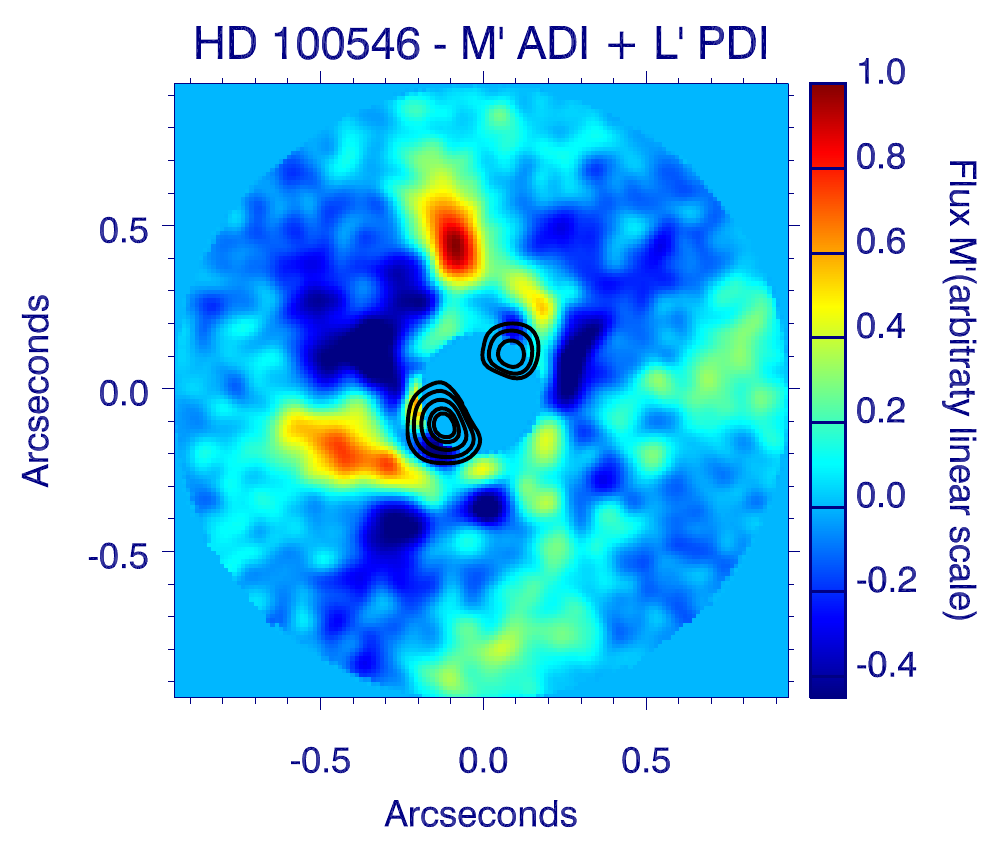}
\caption{Comparison of the extended emission feature southeast of the central star detected in our ADI data with the location of polarized light detected in polarimetric differential imaging (PDI) data. Specifically we compare the $M'$ ADI image (in color), where the extended emission appears to be more pronounced, with the $L'$ PDI image from \citet{avenhaus2014} (contours).\label{fig9}}
\end{figure}

\subsection{The extended emission features}
Finally, concerning the extended emission component in our images, this is likely thermal emission from warm material in the surrounding circumstellar disk. The underlying energy source is unknown at the moment, and whether or not local compressional heating \citep[e.g.,][]{boley2008}, or similar effects, are responsible for the observed flux requires further investigation. As described in section 4.2, deriving flux estimates from our data is not possible without substantial modeling, which is beyond the scope of the current paper. We note that in our final images another region of extended emission is detected to the southeast of the star. It is more clearly seen in the $M'$ images but is also present in $L'$ (see, Figure~\ref{fig1}). There is no point source component included in this emission and its origin is also unclear. Part of the emission, at least in the $L'$ band, might be due to scattered light from the disk surface as in this direction from the star, but at slightly smaller separations, also the $L'$ polarized light images of the disk surface show a flux maximum \citep{avenhaus2014}. We show a direct comparison of the $M'$ image presented here and the polarized light image in Figure~\ref{fig9}. However, given the properties of typical dust grains, the scattering efficiency and hence the detected flux should increase with shorter wavelength. This is difficult to reconcile with our detection in the $M'$ band and no significant emission from this structure in our $K_s$ band images. The feature was also reported by \citet{currie2014} and they propose that it might be a spiral density wave. Indeed, spiral arm features have been reported by various authors and at various locations in the HD100546 disk \citep[e.g.,][]{grady2001,ardila2007, boccaletti2013,avenhaus2014}. More data is required to determine the relative contribution of thermal radiation and scattered light to the observed emission and to understand its physical origin. 
\section{Summary \& Conclusions}
We have presented the first multi-filter study of the protoplanet embedded in the disk of the Herbig Ae/Be star HD100546. Our key results can be summarized as follows
\begin{itemize}

\item The object was detected in $L'$ and $M'$ and consists of an unresolved point source component and a spatially resolved extended emission component. The object was not detected in $K_s$. 

\item The contrast of the point source component relative to the host star is $9.4\pm 0.1$ mag and $9.2 \pm 0.15$ mag in $L'$ and $M'$, respectively. The 3$\sigma$ limit on the minimum contrast in $K_s$ is 9.6 mag. These values translate into apparent magnitudes of $L'=13.92\pm0.10$ mag, $M'=13.33\pm0.16$ mag and $K_s>15.43\pm0.06$ mag.

\item The separation between the point source component and the host star is $(0.457\pm0.014)''$ and $(0.472\pm0.014)''$ in $L'$ and $M'$, respectively. The position angle is $(8.4\pm1.4)^\circ$ and $(9.2\pm1.4)^\circ$. The average de-projected physical separation is $53\pm 2$ au. 

\item Combined with earlier data from 2011 we demonstrated that the object is co-moving with the central star, and also the $L'-M'$ color and apparent magnitudes are inconsistent with any (sub-)stellar fore- or background source. Together with results from other studies our data are best explained with a young forming planet embedded in the HD100546 circumstellar disk.

\item Fitting a single temperature blackbody to the observed fluxes of the point source component yields an effective temperature of $T_{eff}=932^{+193}_{-202}$ K and a radius for the emitting area of $R=6.9^{+2.7}_{-2.9}$ R$_{\rm Jupiter}$. The best-fit luminosity is $L=(2.3^{+0.6}_{-0.4})\cdot10^{-4} L_\sun$. $T_{eff}$ and $R$ increase when possible dust extinction effects caused by the circumstellar environment are taken into account, but they are consistent with the values above at the 1$\sigma$ level.

\item The observed $L'$ and $M'$ magnitudes are inconsistent $(\chi^2\gtrsim25)$ with those predicted by atmospheric models for young (1--10 Myr) gas giant planets. The effective temperature and radii predicted by evolutionary models for young gas giant planets agree with the observations at the $\lesssim$1\% level. The main discrepancy is the large emitting area derived from our data.

\item The large effective emitting area of the object can be readily explained with a combination of a young planet and a surrounding circumplanetary disk. In this case, the derived parameters ($T_{eff}$ and $R$) represent a superposition of the contributions from both components (planet+disk) as the circumplanetary disk is expected to be unresolved in our images.

\end{itemize}

Given these findings, HD100546 is a unique laboratory to study gas giant planet formation empirically. Future ALMA observations with comparable resolution as the data presented here will confirm the existence of the suspected circumplanetary disk and will constrain its extent and mass. Such observations will also yield spatially resolved information about the physical - and potentially chemical - conditions in the circumstellar disk in the vicinity of the forming planet, which may help to further constrain the processes involved in the object's formation. Finally, new high-contrast imaging observations with VLT/SPHERE or Gemini/GPI will further push the detection limits at $K_s$ or even shorter wavelengths and they may even probe directly for the predicted planet orbiting within the disk gap at $\sim$13--14 au \citep{brittain2014}. 

We are entering an era, where we start deriving empirical constraints on the formation sites and formation processes of gas giant planets, and together with HD169142, where also first indications for multiple planet formation have been reported \citep{reggiani2014,osorio2014}, and LkCa15 \citep{kraus2012}, HD100546 will be one of the prime targets for further investigations. 
\\
\acknowledgments

We thank the anonymous referee for a very constructive referee report that improved the initial manuscript. We also thank the staff at Paranal Observatory for their support during the observations. This research made use of the SIMBAD database, operated at CDS, Strasbourg, France, and of NASAÕs Astrophysics Data System. We thank Jaime Pi\~neda, Hans Martin Schmid, Michiel Cottaar and Farzana Meru for useful discussions and Gijs Mulders for providing detailed information about the HD100546 disk model. This work has been carried out within the frame of the National Centre for Competence in Research PlanetS supported by the Swiss National Science Foundation. SPQ and MRM acknowledge the financial support of the SNSF.

{\it Facilities:}  \facility{VLT: Yepun (NACO)}

%\bibliographystyle{apj}
%\bibliography{mybib.bib}

%%\begin{thebibliography}{}

%%\end{thebibliography}{}

\end{document}